\documentclass[aip,apl,amsmath,amssymb,reprint]{revtex4-1}
\usepackage[utf8]{inputenc}

\usepackage{commath}
\usepackage{siunitx}

\usepackage{graphicx}
\usepackage{subfig}

\usepackage[hidelinks]{hyperref}

\usepackage{url}

\DeclareMathOperator{\GammaF}{\Gamma}

\DeclareMathOperator{\dCross}{\sigma}

\newcommand*{\defeq}{\mathrel{\vcenter{\baselineskip0.5ex \lineskiplimit0pt
                     \hbox{\scriptsize.}\hbox{\scriptsize.}}}%
                     =}

\newcommand{\twopartdef}[4]
{
	\left\{
		\begin{array}{ll}
			#1 & \mbox{if } #2 \\
			#3 & \mbox{if } #4
		\end{array}
	\right\}
}

\newcommand{\twopartdefother}[3]
{
	\left\{
		\begin{array}{ll}
			#1 & \mbox{if } #2 \\
			#3 & \mbox{otherwise }
		\end{array}
	\right\}
}

\renewcommand{\eqref}[1]{Eq.~(\ref{#1})}
\newcommand{\beqref}[1]{[Eq.~(\ref{#1})]}

\newcommand{\refcite}[1]{Ref.~\onlinecite{#1}}

\begin{document}

\title{Diagnostic of electron temperature from bremsstrahlung in overdense targets}
\author{G. Hern\'andez}
\author{F. Fern\'andez}
\email{fdz@usal.es}
\affiliation{Universidad de Salamanca, Spain}
\date{\today}

\sisetup{range-units=single}


\begin{abstract}
Models for characterization of laser-accelerated electron via its produced bremsstrahlung are provided for both thin and thick targets. An effective temperature functional is proposed to overcome the so-called cold and hot ``temperatures'' in the emission spectra, which are shown not to describe the underlying electron energy distribution. In contrast, this functional allows for identifying the real effect of a hot electron component. A false ``heating'' effect due to added noise is also exposed. The models are shown to be in good agreement with Monte Carlo simulations, as well as in good agreement with other experimental methods when applied to experimental data.
\end{abstract}

\maketitle

\section{Introduction}\label{sec:introduction}
The rapid development of short pulse ultraintense lasers has led to extensive investigation in laser driven plasmas as a source of high energy charged particles. Even table top laser systems allow the direct acceleration of electrons up to MeV energies when focused on solid targets.\cite{malka:1996}
The random nature of the particle acceleration leads to strong cycle-to-cycle fluctuations in the electron trajectories and energies, although the average of these single particle distribution leads to Maxwellian energy distributions \cite{bezzerides:1982,mordovanakis:2010} which are usually characterized by a temperature-like parameter.

Some works in the literature\cite{estabrook:1978} also describe these distributions as bi-Maxwellians, characterized by two temperature parameters plus a mix parameter. The lower energy particles would correspond to the initial thermal distribution of the preformed plasma electrons, whereas the higher energy distribution represents the electrons heated trough collisionless absorption mechanisms (resonance absorption, vacuum heating, $\mathbf{J} \times \mathbf{B}$ heating, etc\cite{wilks:1997}).

When these electrons interact with surrounding material, they produce x-ray pulses which spectrally consist of a continuous bremsstrahlung component and characteristic lines emission. The pulse duration (of the order of few hundreds of femtoseconds\cite{radunsky:2007}) and the source size (few times larger than the laser spot size) make these x-rays sources very useful for many applications like time resolved diffraction, medical imaging, spectroscopy, and microscopy of transient phenomena.

The direct determination of the temperature of a certain electron distribution requires an extensive measurement, because this parameter is to be obtained asymptotically from a log-linear plot of the energy distribution. This measurement is even more complicated when the experiments are performed in air.

For that reason, and taking into account that the x-ray energy spectra should be related with those of the electrons, another proposal also found in the literature is to measure the `temperature' of the x-ray spectra, easier to obtain with solid-state detectors, to try to characterize the two Maxwellian distributions of the electron population by fitting two temperatures in two different regions of the spectra.\cite{zulick:2013} However, as we shall show in this work, the relationship between the electron temperatures and the parameters that can be extracted --when the signal-noise ratio allows it--- from the x-ray spectra is not straightforward.

An alternative, more rigorous method would be to simulate, typically by Monte Carlo methods, the bremsstrahlung production for a set of distributions and take the least-squares best fit of the outputs, with the disadvantage of the computational cost.\cite{zulick:2013}

In this work we will propose another method to address this problem, providing some models for thin and thick target with different refinement will be proposed to relate Maxwellian electron distributions and their resulting bremsstrahlung. A method of identifying the effect of a two-temperature electron distribution will also be shown. These models have some advantages over the simulation methods described before, like allowing correcting existing log-scale temperature fits, decreasing the needed computation time, and revealing the physical effects responsible for the difference between electron and photon distributions.

\section{Method}\label{sec:method}

\subsection{The notion of effective temperature}\label{sec:methodMeaning}

\begin{figure}[!ht]
\centering
\includegraphics[width=0.48\textwidth]{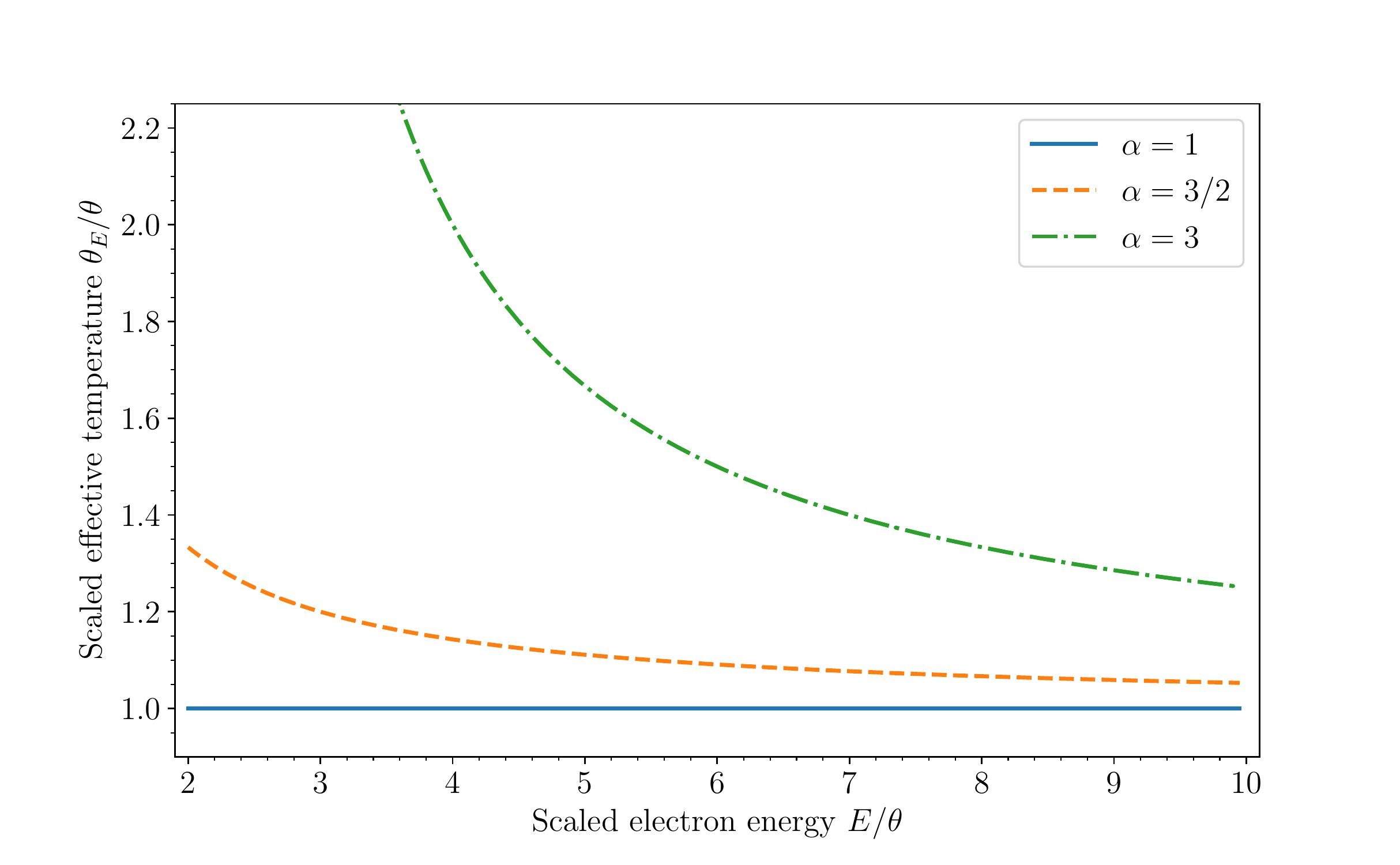}
\caption{\label{fig:EffTempGamma}Effective temperature in gamma distributions as a function of energy, both scaled by the temperature $\theta$, for different values of the shape parameter $\alpha$.}
\end{figure}

The ``Maxwellian'' electron energy distributions generated by laser interaction in overdense plasmas can all be described using the general form of the probability density function of a gamma distribution,
\begin{equation}\label{eq:gamma}
 f_{\alpha,\theta}(E)=\frac{E^{\alpha-1}}{\Gamma \left( \alpha \right) \theta^\alpha} \mathrm{e}^{-E/\theta} \,,
\end{equation}
where $E\in \left[0, \infty \right)$ is the electron kinetic energy, $\theta$ is a scale parameter with dimension of energy which is usually called `temperature', and $\alpha \in \left(0, \infty \right)$, usually an integer or a semi-integer, is a shape parameter which, in some physical systems, is identified with half the number of degrees of freedom. Both the terms ``Maxwellian'' and ``gamma'' will be used to refer to these distributions. Typical examples include\cite{batani:2010} the exponential distribution ($\alpha=1$), the Maxwell-Boltzmann distribution ($\alpha=3/2$), and the ultrarelativistic Maxwell-J\"uttner distribution ($\alpha=3$); as well as other forms sometimes introduced for analytical simplicity (e.g., $\alpha=2$ was introduced in \refcite{Galy:2007}).
It is a well known fact that the slope of this distributions in a log-linear plot can be used to estimate $\theta$. This can be shown by considering the logarithmic derivative of the distribution function,
\begin{equation}\label{eq:dlog}
 \od{\ln{f_{\alpha,\theta}}}{E}= \frac{f'_{\alpha,\theta}(E)}{f_{\alpha,\theta}(E)}= \frac{\alpha-1}{E}-\frac{1}{\theta} \,,
\end{equation}
which clearly tends to $-\theta^{-1}$. It follows that a temperature being estimated from any energy distribution $f$ using energies in a neighborhood of $E$ is indeed given by a functional
\begin{equation}\label{eq:effTemp}
 \theta_E\left[f\right] \defeq -\frac{f(E)}{f'(E)}\,.
\end{equation}
We shall call this functional effective temperature. For gamma distributions we have
\begin{equation}\label{eq:effTempGamma}
 \theta_E\left[f_{\alpha,\theta}\right]=\frac{E/\theta}{E/\theta + \left(1-\alpha\right)}\theta \,.
\end{equation}
The behavior of \eqref{eq:effTempGamma} is depicted in~\figref{fig:EffTempGamma}. It is worth noting that accurate estimation of $\theta$ relies on the asymptotic part of the spectrum being used in the fit, unless the distribution is an exponential. Otherwise, overestimation of the temperature will occur unless the asymptotic condition $E\gg\theta$ holds.

In real fittings estimations in a neighborhood are not possible and intervals $\left[E_1,E_2\right]$ are used instead. In this case $\ln{f(E)}$ is being fit to a line
\begin{equation*}
a_{E_1,E_2}-E/\theta_{E_1,E_2}
\end{equation*}
in the sense of least squares, where $a$ and $\theta$ are parameters. In the limit case of uniform sampling this is equivalent to the well known problem of find the best 1-degree polynomial fit in $L^2[E_1,E_2]$. From the solution of that problem~\cite{hildebrand:1987} it is immediate to state

\begin{equation}
\begin{split}
\label{eq:effTempInt}
 &\theta_{E_1,E_2}\left[f\right]=\\
 &\frac{\left(E_2-E_1\right)^3}
 {12 \int_{E_1}^{E_2} E \ln f(E) \dif E  - 6\left(E_2+E_1\right) \int_{E_1}^{E_2} \ln f(E) \dif E   }
\end{split}
 \,.
\end{equation}

A simple analytical solution also exists for these gamma distributions, given by
\begin{equation}\label{eq:effTempIntGamma}
 \theta_{E_1,E_2}\left[f_{\alpha,\theta}\right]= 
 \frac{\frac{(E_2-E_1)^3}{\theta^3}}
 {3 (\alpha -1) \frac{ 2E_2 E_1 \ln \left(\frac{E_2}{E_1}\right)-\left(E_2^2-E_1^2\right)}{\theta^2}+\frac{(E_2-E_1)^3}{\theta^3}}\theta \,.
\end{equation}

\subsection{Bremsstrahlung calculations}\label{sec:methodBremss}
There are two simple hypothesis on the medium where the accelerated electrons produce the bremsstrahlung. On the one hand, the amount of material the electrons travel through can be viewed as negligible (``thin target''); for example, a narrow cone of electrons emitted from the surface of the plasma to the outside of the target. On the other hand, the amount of material can be regarded as big enough for almost all electrons to stop completely (``thick target''); for example, for isotropic electrons emitted from the plasma, as long as the dimensions of the target are bigger than the range of most of the electrons in the distribution. The continuous slowing down approximation range can be used for this purpose.\cite{star:2009}

\subsubsection{Thin target models}
The bremsstrahlung emitted by a general electron energy distribution $f$ is given by its integral with the angle-integrated production cross-section, i.e.,
\begin{equation}\label{eq:bremssIntegral}
 f_\gamma(E_\gamma) \propto \int_{E_\gamma}^\infty f(E) \dCross{\left(E,E_\gamma\right)} \dif E \,,
\end{equation}
where $\dCross$ is the angle-integrated bremsstrahlung cross-section. In the thin target approximation changes in the electron energy and the photon attenuation along the material are both neglected and the emission distribution~\eqref{eq:bremssIntegral} is assumed to be the detected distribution.

The general dependence of the angle-integrated bremsstrahlung cross-section $\dCross$ with the electron energy can be described by the almost linear dependence of the scaled cross-section $E_\gamma \dCross$ with the scaled energy $E_\gamma/E$, which suggest an approximation of the form
\begin{equation}\label{eq:CSModel}
\dCross{\left(E,E_\gamma\right)} \propto \twopartdefother{E_\gamma^{-1}-b E^{-1}}{E > E_\gamma}{0}\,,
\end{equation}
where $b\leq1$ is a dimensionless positive constant that must be fitted to experimental values of the cross-section. The condition in the definition guarantees only non-negative values are predicted. Such an approximation was studied by Findlay in~\refcite{findlay:1989}, where the value $b=0.83$ was claimed to accurately represent the results of~\refcite{seltzer:1986} in the \SIrange{5}{20}{\MeV} electron kinetic energy range.

Other specific fits on $b$ can be attempted in order to improve the cross-section description in a different energy range. Numerical integration of \eqref{eq:bremssIntegral} with a varying upper integration limit can be used to show that, for a fixed $E_\gamma$, most of its production comes from electron energies about two or three times its value. Thus, given an electron energy $E$ in the range of  interest, it is more important to reproduce the lower energy part of the $\dCross(E,E_\gamma)$ than its ``tip''. The value $b=1$, which corresponds to neglecting the tip, can be checked to reproduce this part of the cross-section accurately. There also exists another advantage in favor of this model that will be discussed later in this section. However, note that for different purposes other values of $b$ might provide a better description, specially when then bremsstrahlung tip is relevant.

By using \eqref{eq:bremssIntegral} and \eqref{eq:CSModel}, analytical predictions of the bremsstrahlung spectra are possible, under the approximations of thin target and Findlay's cross-section. In the case of gamma distributions of electrons this can be expressed as
\begin{equation}\label{eq:bremssIntegralGamma}
\begin{split}
f_{\gamma;\alpha,\theta}^{\textrm{Fin}; b}(E_\gamma) & \propto \int_{E_\gamma}^\infty f_{\alpha,\theta}(E) (E_\gamma^{-1}-b E^{-1}) \dif E \\
 & = \frac{\theta  \GammaF \left(\alpha ,\frac{E_\gamma}{\theta }\right)-b E_\gamma \GammaF \left(\alpha -1,\frac{E_\gamma}{\theta }\right)}{E_\gamma \theta  \GammaF (\alpha )}
\end{split}
\,,
\end{equation}
where $\GammaF{\left(a,z\right)}$ is the incomplete gamma function as defined in~\refcite{Olver:2010:NHMF,NIST:DLMF}.

The effective temperature can be calculated by means of~\eqref{eq:effTemp}, yielding
\begin{equation}\label{eq:EffTempGammaBremss}
 \theta_{E_\gamma}\left[f_{\gamma;\alpha,\theta}^{\textrm{Fin}; b}\right]= 
 \frac{   \frac{E_\gamma}{\theta} \GammaF{\left(\alpha ,\frac{E_\gamma}{\theta }\right)}-b \frac{E_\gamma^2}{\theta^2} \GammaF{\left(\alpha -1,\frac{E_\gamma}{\theta }\right)}}
 {\left((1-b) e^{-E_\gamma/\theta } \left(\frac{E_\gamma}{\theta }\right)^{\alpha }+ \GammaF{\left(\alpha ,\frac{E_\gamma}{\theta }\right)}\right)}
 \theta
 \,.
\end{equation}
The asymptotic behavior is not as obvious as it was in~\eqref{eq:effTempGamma}, but making use of the asymptotic expansion of the incomplete gamma function (\S8.11(i) in \refcite{NIST:DLMF}) one can find the series expansion
\begin{equation}\label{eq:EffTempGammaBremssLimit}
 \theta_{E_\gamma}\left[f_{\gamma;\alpha,\theta}^{\textrm{Fin}; b}\right]= 
 \theta \left(1 + \left(\alpha -\twopartdef{2}{b<1}{3}{b=1}\right) \frac{\theta}{E_\gamma}+ \mathcal{O}\left(\frac{\theta^2}{E_\gamma^2}\right) \right)
 \,.
\end{equation}

\begin{figure}[!ht]
\centering
\includegraphics[width=0.45\textwidth]{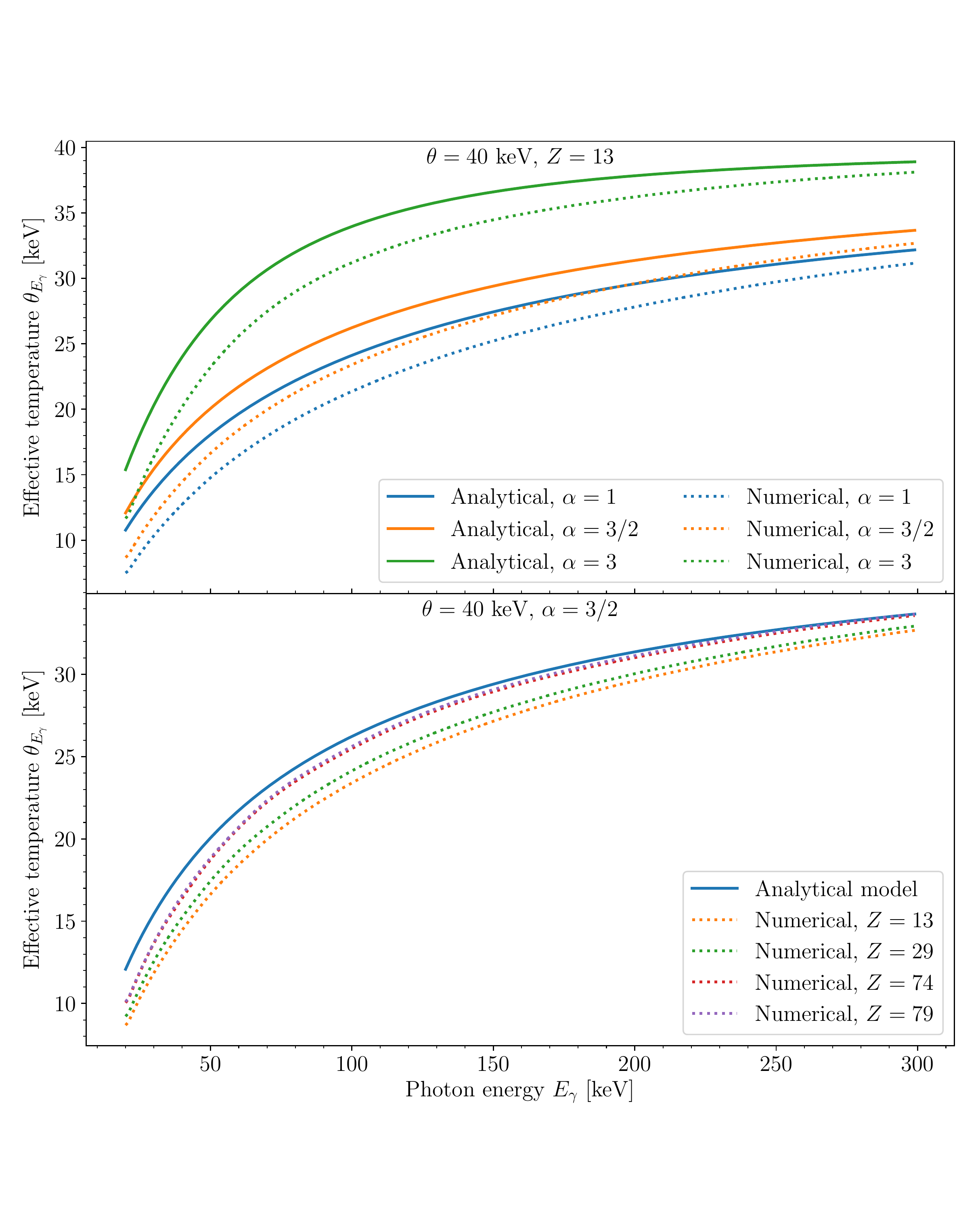}
\caption{\label{fig:EffTempThin}Effective temperature in a gamma distributed electron-produced bremsstrahlung as a function of the photon energy for different values of the shape parameter $\alpha$ and material $Z$. Solid lines show the model given in \eqref{eq:EffTempGammaBremss}, dotted lines show the numerical calculations of \eqref{eq:bremssIntegral} using tabulations derived from the works of \refcite{seltzer:1986}. The plot above shows calculations a low $Z$ material (Al), where the maximum deviation from the model was found. The plot below shows the change of the functional with different $Z$ values. The results are similar for different temperatures in the same range.}
\end{figure}

Despite it is still true that the effective temperature functional can be used to estimate the temperature in the asymptotic limit $E_\gamma \gg \theta$, in practice the condition worsens, as it can be seen in \figref{fig:EffTempThin}, where the exact result of ~\eqref{eq:EffTempGammaBremss} has been depicted for $b=1$. Direct numerical calculations using \eqref{eq:bremssIntegral} with data from the Seltzer and Berger description\cite{seltzer:1986} are also shown as dashed lines. It is worth noting that, assuming $b=1$, in contrast to what was seen in~\eqref{eq:effTempGamma} and \figref{fig:EffTempGamma} the temperature is underestimated if $\alpha\leq3$ (the $\alpha=3$ case predicts a second-order underestimation), so the exponential-generated bremsstrahlung is not exempt from this effect, and both overestimation and underestimation might occur if $\alpha>3$, depending on the energy region being used. Alternative modeling with $b<1$ fails to reproduce this behavior which is numerically observed in \figref{fig:EffTempThin}, which is another reason supporting the choice made.

The deviation found with high Z materials like tungsten or gold is below \SI{2}{\percent} for $E_\gamma\gtrsim\theta$ and thus the model can be applied safely in the keV energy range. For low Z materials like copper or aluminum this increases to around \SI{10}{\percent}. In these cases numerical calculations should be preferred.

An analogous expression for the calculation on an interval can be obtained from~\eqref{eq:effTempInt}, which requires numerical calculations.

\subsubsection{Thick target models}
As introduced before, the alternative hypothesis to the thin target is that the thickness is big enough so all electrons stop on it.

A simple approximation to the emission from a thick target can be obtained assuming the intensity emitted from a target follows a linear model reaching zero at the energy limit given by the Duane-Hunt law (cf. p. 272 in \refcite{podgorsak:2010}), the so-called Kramer's model.\cite{kramers:1923} Thus, the number of photons emitted by an electron of energy $E$ are roughly  given by
\begin{equation}
\sigma_{\textrm{Kra}}(E, E_\gamma) \propto \twopartdefother{\frac{E_e}{E_{\gamma }}-1}{E_{\gamma }\leq E_e}{0}\,.
\end{equation}
This expression behaves like a cross-section in the sense that its substitution in \eqref{eq:bremssIntegral} yields the total bremsstrahlung emitted from the target. From this follows that
\begin{equation}
f_{\gamma;\alpha,\theta}^{\textrm{Kra}} \propto
\frac{\theta  \GammaF{\left(\alpha +1,\frac{E_{\gamma }}{\theta }\right)} - E_{\gamma } \GammaF \left(\alpha ,\frac{E_{\gamma }}{\theta }\right)}{E_{\gamma } \Gamma (\alpha )}
\end{equation}
and
\begin{equation}\label{eq:tempKramers}
 \theta_{E_\gamma}\left[f_{\gamma;\alpha,\theta}^{\textrm{Kra}}\right] =
 \left(\frac{E_{\gamma }}{\theta}-\frac{E_{\gamma }^2}{\theta^2}\frac{ \GammaF \left(\alpha ,\frac{E_{\gamma }}{\theta}\right)}{\GammaF \left(\alpha +1,\frac{E_{\gamma }}{\theta }\right)} \right) \theta
 \,.
\end{equation}

It is worth noting, comparing \eqref{eq:tempKramers} and \eqref{eq:EffTempGammaBremss}, that 
\begin{equation}\label{eq:KramersAndThin}
\theta_{E_\gamma}\left[f_{\gamma;\alpha,\theta}^{\textrm{Kra}}\right] = 
\theta_{E_\gamma}\left[f_{\gamma;\alpha+1,\theta}^{\textrm{Fin}; 1}\right]\,.
\end{equation}
This means that difference in the shape of a bremsstrahlung spectra due to the medium being thick instead of thin is equivalent to that of an increase in a ``degree of freedom'' of the electron population. Hence, these two effects are difficult to identify by simply analyzing an spectrum.

A more comprehensive physical description of thick target bremsstrahlung can be obtained by characterizing the electron fluence in the target, integrating a cross section, and taking into account the intrinsic attenuation of the medium. To this purpose we will use the model of \refcite{hernandez:2016}, for which an implementation (xpecgen) is also available.\cite{hernandez:2016b}

\subsubsection{Comparison with finite material simulations}\label{sec:results:numerical}
To study the effect of the finite amount of material where the bremsstrahlung is generated some simulations were performed with the the Monte Carlo package FLUKA 2011.2c.5~\cite{bohlen:2014,ferrari:2005}. A conical beam with \SI{17}{\degree} divergence of Maxwellian electrons of different temperatures was simulated to cross a 1 or \SI{10}{\micro m}-thick tungsten region. Each simulation consisted of a number of primaries between \SI{1.8E8}{} and \SI{3.5E9}{}, depending on the simulation. Single scattering was activated in all regions for all charged particles and production and transport thresholds were set at $\SI{1}{\keV}$ for electrons and photons. The bremsstrahlung decay length was biased by a factor of 0.2, which is properly account in the particle weights.

\begin{figure}[!ht]
\centering
\includegraphics[width=0.48\textwidth]{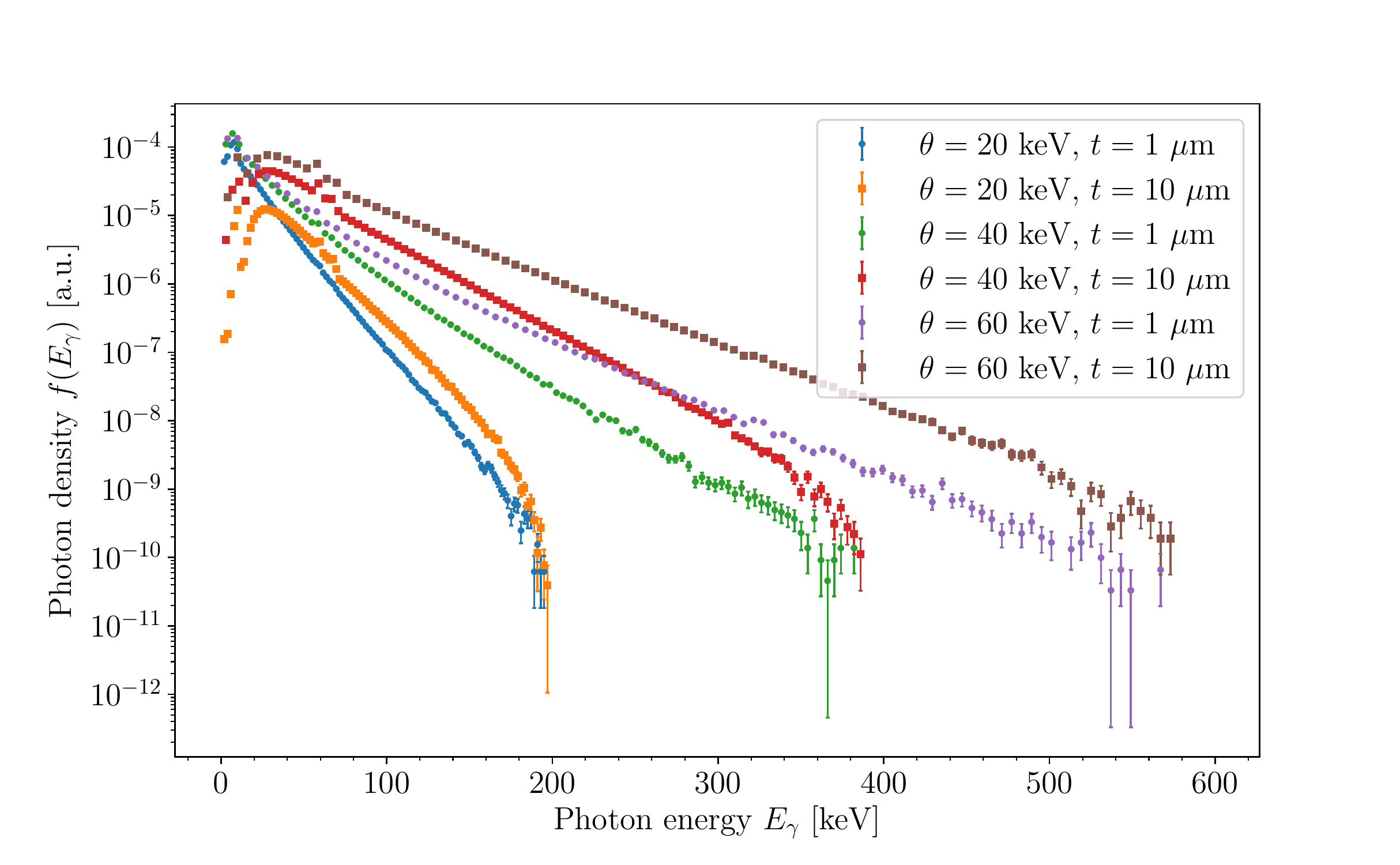}
\caption{\label{fig:SpectraFLUKA}Spectra produced in a finite-thickness tungsten converter calculated with FLUKA.}
\end{figure}

\begin{figure}[!ht]
\centering
\includegraphics[width=0.45\textwidth]{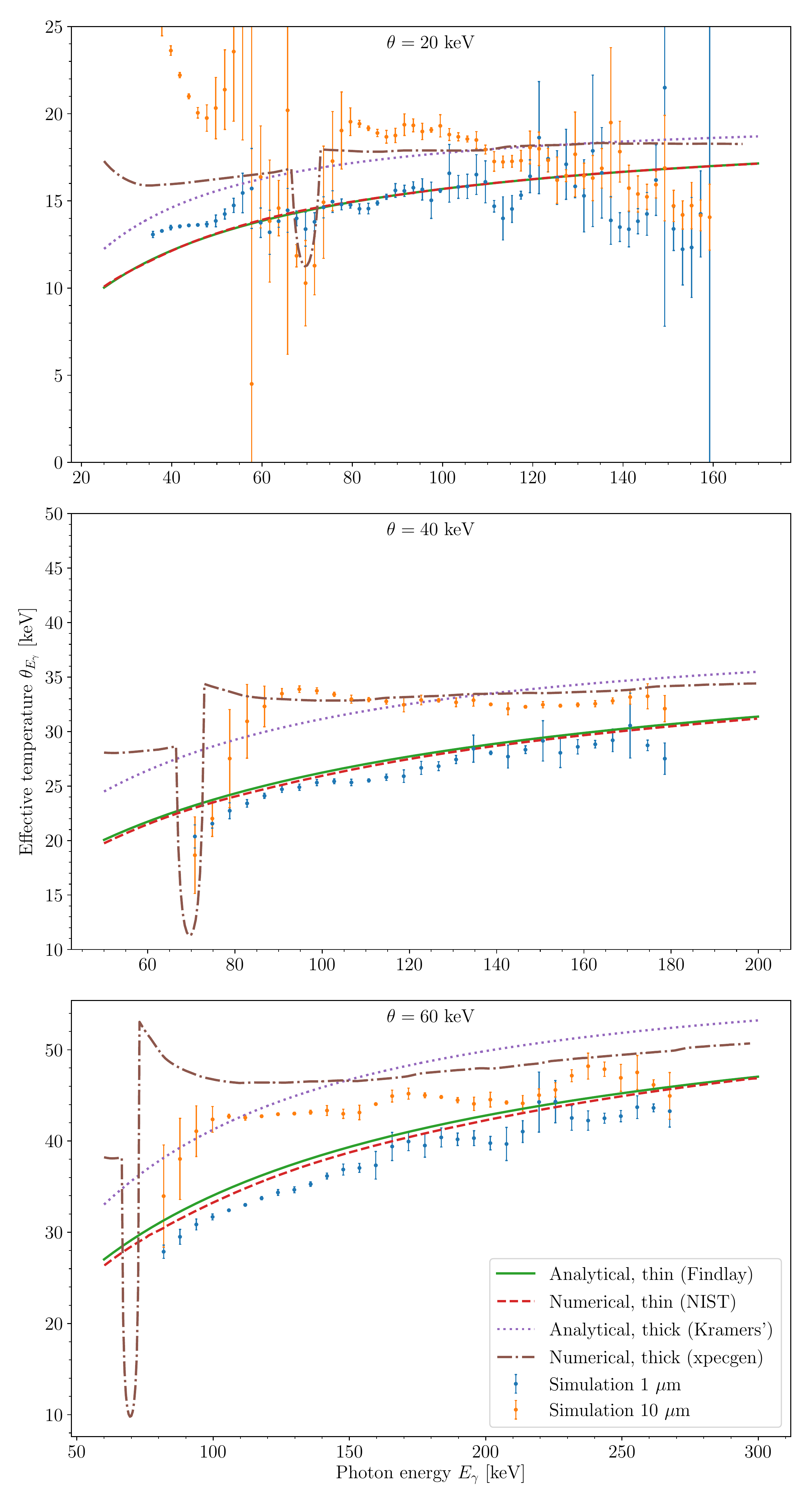}
\caption{\label{fig:EffTempFLUKA}Comparisons of the models (lines) with finite depth simulations done with FLUKA (points). Solid lines show the model given in \eqref{eq:EffTempGammaBremss}, dashed lines show the numerical calculations of \eqref{eq:bremssIntegral} using tabulations derived from the works of \refcite{seltzer:1986}, dotted lines show the analytical thick model of \eqref{eq:tempKramers}, and dot-dashed lines the numerical calculations using \refcite{hernandez:2016b}.}
\end{figure}

The simulated spectra are shown in \figref{fig:SpectraFLUKA}. From this data, effective temperatures were calculated by fitting sets of up to 10 sampling points. These are shown in \figref{fig:EffTempFLUKA}, with error bars depicting the standard deviation of the fit value when the number of points is reduced. The predictions of both thin and thick target models are also shown in \figref{fig:EffTempFLUKA}.

In general, the $\SI{1}{\micro m}$ simulations match both of the thin target models. With respect to the $\SI{10}{\micro m}$ case, both of the thick target models globally reproduce the simulated effective temperatures for $\theta = \SI{20}{keV}, \SI{40}{keV}$. For the \SI{60}{keV} case, the simulation results lie between the predictions of thin and thick target models. This could be expected since, for high $Z$ targets, the penetration depth is around $2r_\text{CSDA}$ and $r_\text{CSDA}(\SI{50}{keV})\approx \SI{5}{\micro m}$. Thus, for energies is this range a target with such a thickness can be regarded neither as thin nor thick.

Note the abrupt region around $\SI{70}{keV}$ where the temperature decreases is related with the K-alpha absorption edge. The numerical thick model (xpecgen) is the only one to take into account the intrinsic attenuation of the material, and thus, the only one to reproduce it. Once again, one can see that the effective temperature obtained from the x-ray spectrum underestimates the true electron temperature.

\subsection{Mixtures of distributions}\label{sec:methodMixtures}
\subsubsection{Two-temperature electron populations}\label{sec:TwoTemperatures}
As stated in the introduction, it is usual in the literature to explain x-ray spectra as being produced from mixtures of ``cold'' and ``hot'' electrons (a ``bi-Maxwellian'' distribution), so the changes in the log-scale slope found in the spectra seem explained. However, as we have previously seen, this behavior is also characteristic of single-temperature-produced bremsstrahlung. A specific process for identifying electron temperature mixtures is thus needed.

From the effective temperature functional~\beqref{eq:effTemp}, applying simple algebra one can find that, for a convex combination of distributions with PDFs $f_i$ and weights $1\geq a_i\geq 0$ such that $\sum a_i = 1$,
\begin{equation}\label{eq:convexCombination}
 f(E)=\sum_{i}{a_i f_i(E)}\,,
\end{equation}
the effective temperature can be written as
\begin{equation}\label{eq:convexCombinationTemperature}
 \frac{1}{\theta_E\left[f\right]}=\sum_{i}{\frac{a_i f_i(E)}{f(E)} \frac{1}{\theta_E\left[f_i\right]}}\,,
\end{equation}
which has a clear meaning: the inverse of the effective temperature is a convex combination of the individual ones with weights that depend on the relative amount of their PDFs.

Hence, if one wants to explain an effective temperature around an energy $E$ using a single component component of the mixture $a_i f_i(E) \gg a_j f_j(E)$ must hold for every $j \ne i$, or otherwise both components must be taken into account to make a prediction.

\begin{figure*}[ht]
\centering
\includegraphics[width=0.90\textwidth]{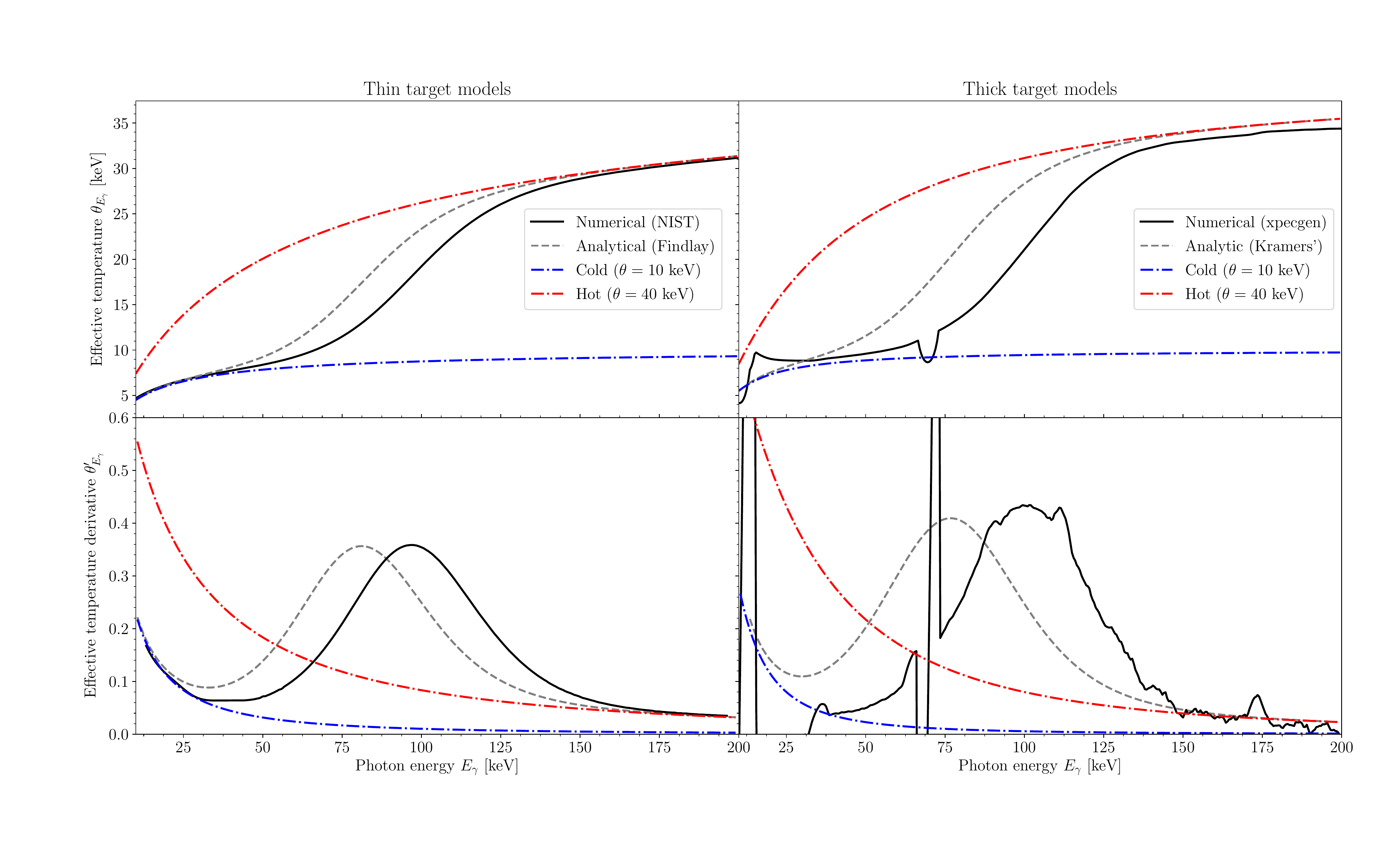}
\caption{\label{fig:EffTempMixtureBoth}Effective temperature (above) and its derivative (below) in the bremsstrahlung produced by a bi-Maxwellian electron distribution using the cross-section model \beqref{eq:CSModel} (solid line), numerical calculations from the tabulations from the works of \refcite{seltzer:1986} (dashed lines), and monoenergetic models \beqref{eq:EffTempGammaBremss} (dot-dashed lines). The hot component fraction is \SI{0.5}{\%}.}
\end{figure*}

The typical behavior of the effective temperature functional for bi-Maxwellian electron-produced bremsstrahlung, using the different models, is depicted in \figref{fig:EffTempMixtureBoth}, where $\theta_{\textrm{cold}}=\SI{10}{keV}$, $\theta_{\textrm{hot}}=\SI{40}{keV}$, and the hot electron fraction is \SI{0.5}{\%}. Both single temperature Maxwellian cases are also shown there. For small photon energies the cold electron component dominates, while the high energy tail follows the hot component instead. In between these two situations both of them are relevant to explain the transition, where the functional experiences a change in convexity in a energy interval, easier to identify in its derivative, also depicted in the figure.

\subsubsection{Noise influence in the effective temperature}\label{sec:noise}
\begin{figure}[ht]
\centering
\includegraphics[width=0.45\textwidth]{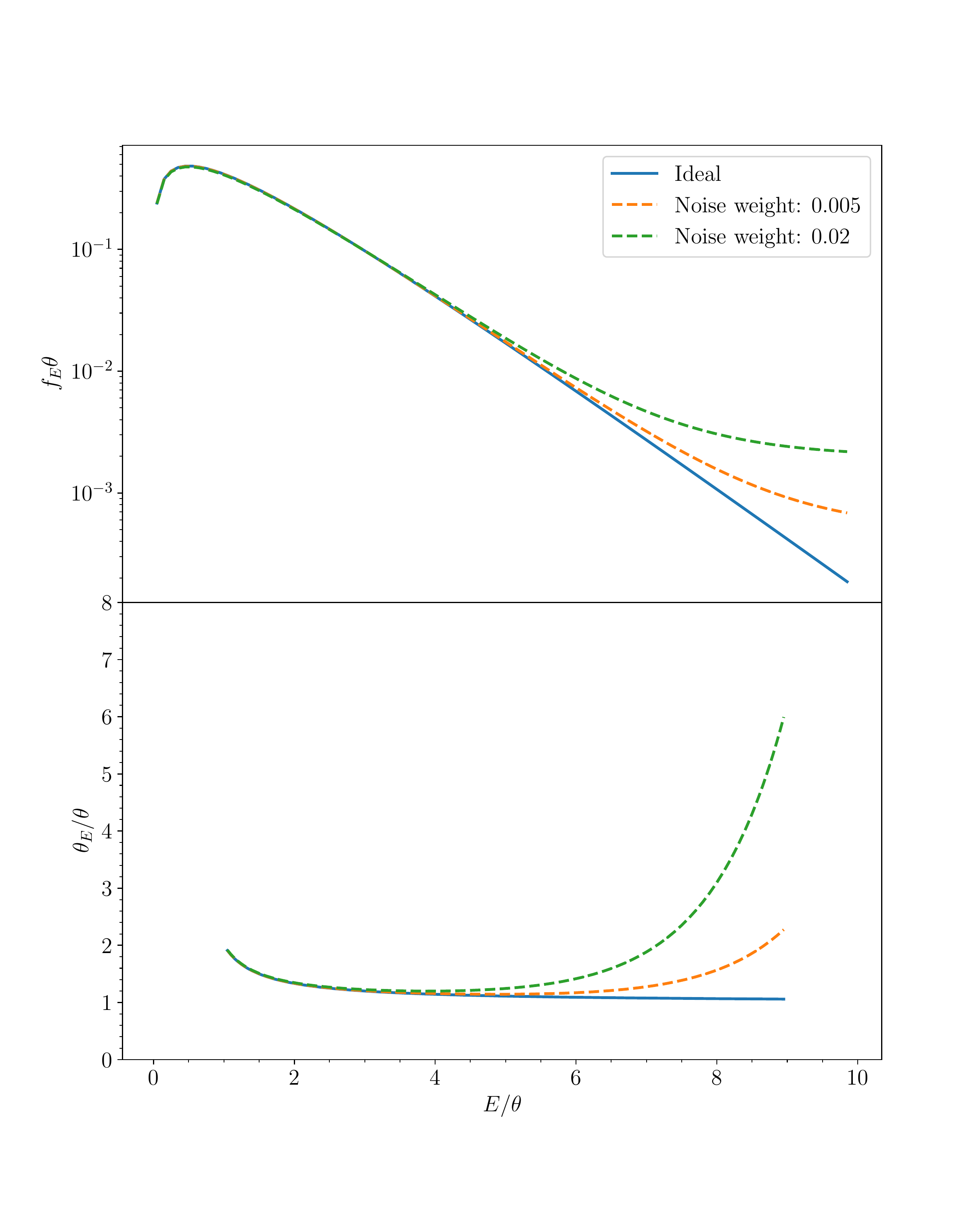}
\caption{\label{fig:Noise}Impact of an added uniform noise distribution in a Maxwellian distribution. The change in the slope of the spectrum could be though as a second temperature, but the shape of the effective temperature functional makes clear it is not so (cf.~\figref{fig:EffTempMixtureBoth}).}
\end{figure}

As we have shown above, electron temperatures are only found directly in their bremsstrahlung spectra in the asymptotic limit. However, in the high energy region of the spectra the number of events is usually small, and random noise can be important. Assuming a uniform noise is present in the detector, it will behaves like a distribution whose inverse temperature is zero. From \eqref{eq:convexCombinationTemperature} follows that, when this noise is relevant, the effective temperature increases in a seemingly unphysical way, since the resulting effective temperature cannot be attributed to an electron population generating its bremsstrahlung. An example with Maxwellian models, chosen for the sake of simplicity, is shown in \figref{fig:Noise}, where this Maxwellian spectra with different degrees of noise are plotted (panel above), as well as the effective temperature calculated using \eqref{eq:convexCombinationTemperature} and \eqref{eq:effTempGamma}. This situation should be contrasted with the two-temperature shown in \figref{fig:EffTempMixtureBoth}. The effective temperature functional tends to diverge as the noise increases, while a true hot electron component will asymptotically approach a finite value. Plotting of the functional would serve to clarify the issue.

\section{Results and discussion}\label{sec:results}
\subsection{Experimental application}\label{sec:results:experimental}

\begin{figure}[!ht]
\centering
\includegraphics[width=0.45\textwidth]{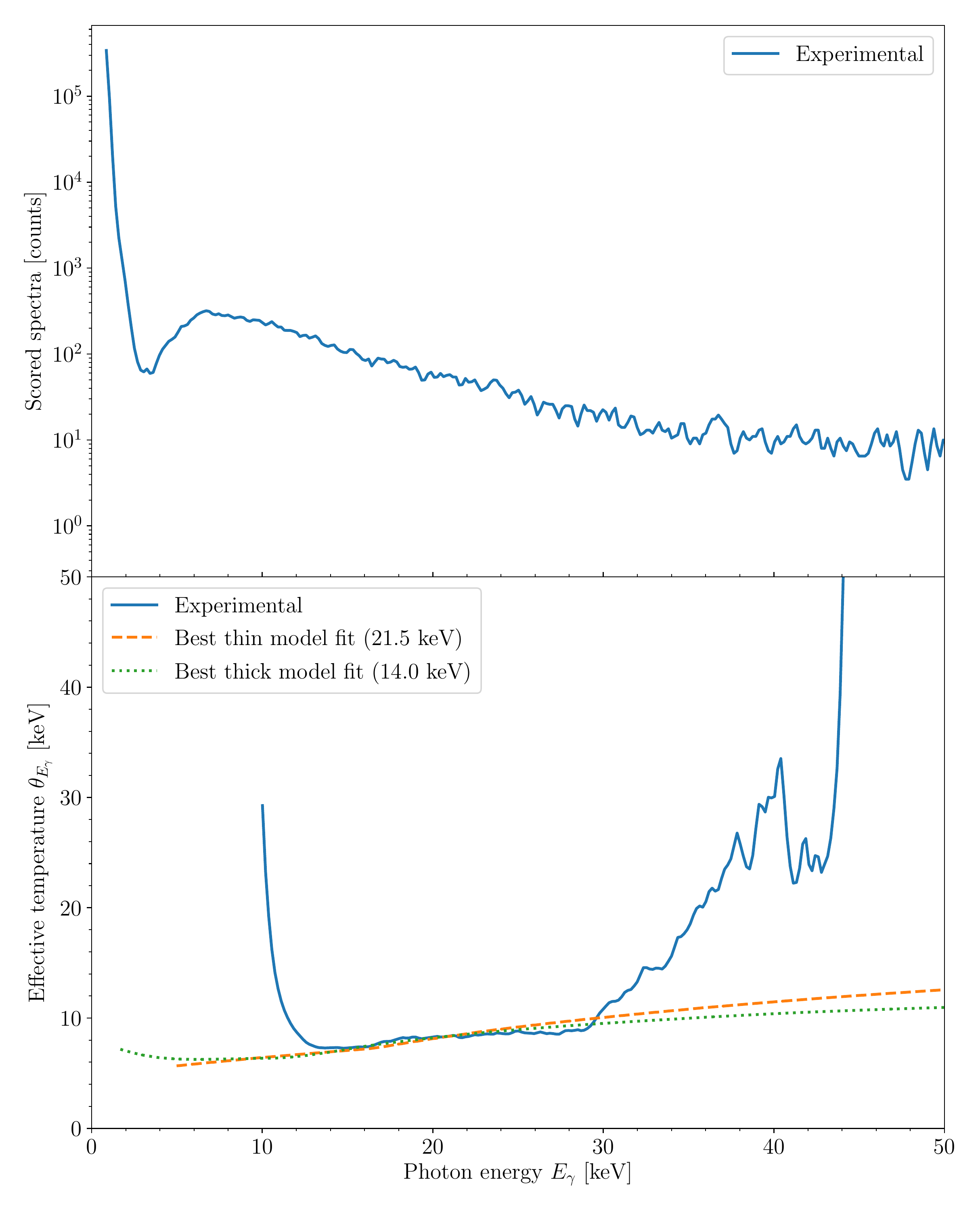}
\caption{\label{fig:EffTempCarmen}Comparisons of experimental data (solid line) with a best fit of the numerical thin target model (numerical calculation using tabulations from \refcite{seltzer:1986}, dashed line) and with numerical thick target model (derived from xpecgen, dotted line).}
\end{figure}

A comparison with the experimental data from \refcite{fonseca:2011} is given below. This experiment was performed with a \SI{1}{GW}, \SI{990}{Hz} Ti:Sa laser, focused with an estimated intensity of $\SI{5.4E16}{W cm^{-2}}$ in an Aluminum target. The bremsstrahlung, measured with an Amptek XR-100T-CdTe, is shown in \figref{fig:EffTempCarmen}, as well as the effective temperature functional. The models best fitting the region where the effective temperature varies smoothly, $E_\gamma \in \left[15, 25 \right] \SI{}{keV}$, are also shown as the dashed line (thin target) and as the dotted line (thick target). Only analytical models were considered here, since the experiment was not performed with a high Z target. The thick target model, which is the reasonable choice assuming an approximately isotropic electron distribution is being formed in the target, predicts an electron temperature of $\theta\approx\SI{14.0}{keV}$. This is in remarkable agreement with the results of \refcite{fonseca:2011}, where $\theta\approx\SI{13.8}{keV}$ was claimed to explain the dose-distance curves found with a TLD detector.

The abrupt increase in the effective temperature beyond \SI{30}{keV} is not explained by a single temperature, and beyond \SI{45}{keV} the values are too big to be physical (hundreds of keV), which can be only attributed to noise as discussed in \secref{sec:noise}. While there could be a additional hot electron component in the source, it cannot be distinguished in the photon spectra, because the asymptotic condition it would need to manifest is absent.

\section{Conclusions}\label{sec:conclusions}
Direct identification of the slope of the log-linear plot of a photon distribution with the electron source temperature has been shown to underestimate such a parameter. Furthermore, mixtures of two Maxwellian electron distribution cannot be identified in general by fitting two regions of the spectra, and, if doing so, noise can be mistaken with a false hot electron component.

Analytical and numerical models have been developed to characterize the Maxwellian electron-produced bremsstrahlung both in thin and thick targets. They allow to extract the electron characteristics using the effective temperature functional, which behaves like a moving-window fit in log-scale of the spectra. These models also provide a characterization of bi-Maxwellian electron-produced bremsstrahlung which can be distinguished from single temperature effects and from noise influence on the high energy tail of the spectra.

The models were shown to be in good agreement with simulated and experimental data and can be used to estimate the underlying electron distribution parameters.

\section*{Acknowledgments}
One of the authors (G. H.) gratefully acknowledges the Consejer\'ia de Educaci\'on de la Junta de Castilla y Le\'on and the European Social Fund for financial support.

\bibliographystyle{vancouver}
\bibliography{temperature}

\end{document}